\documentclass[12pt]{article}
\usepackage{geometry}
\usepackage[utf8]{inputenc}
\PassOptionsToPackage{hyphens}{url}
\usepackage{hyperref}

\usepackage{booktabs}       
\usepackage{amsfonts}       
\usepackage{nicefrac}       
\usepackage{microtype}      

\usepackage{graphicx}
\usepackage{subfigure}
\usepackage{enumitem}
\usepackage{url}            
\usepackage{braket} 
\usepackage{amsmath}
\usepackage{color}
\usepackage{mathtools}
\usepackage[toc,page]{appendix}
\usepackage{amsthm}
\usepackage{wrapfig}
\usepackage{float}
\usepackage{tcolorbox}
\usepackage{thmtools}
\usepackage{thm-restate}
\newtheorem{theorem}{Theorem}
\usepackage{algpseudocode}
\usepackage[ruled,vlined]{algorithm2e}

\usepackage{url}
\urldef\myurlevalmode\url{https://louisabraham.github.io/crackovid/crackovid.html?input=MyAzCgowLjk5IDAuOTUKCjAuMSAwLjEgMC4xIAoKZXZhbAoKMDExCjEwMQoxMTAKCjAwMA%3D%3D}

\urldef\myurloptimmode\url{https://louisabraham.github.io/crackovid/crackovid.html?input=MyAzCgowLjk5IDAuOTUKCjAuMSAwLjEgMC4xCgpvcHRpbSBjb25maWRlbmNlCmdhLWx1YnkgMiAxMDAKMTAwMA%3D%3D}

\title{Crackovid: Optimizing Group Testing}
\author{Louis Abraham, Gary B\'ecigneul, Bernhard Sch\"olkopf }
\date{April 2020}

\begin{document}

\maketitle

\begin{abstract}
    We study the problem usually referred to as \textit{group testing} in the context of COVID-19. Given $n$ samples taken from patients, how should we select mixtures of samples to be tested, so as to maximize information and minimize the number of tests? We consider both adaptive and non-adaptive strategies, and take a Bayesian approach with a prior both for infection of patients and test errors. We start by proposing a mathematically principled objective, grounded in information theory. We then optimize non-adaptive optimization strategies using genetic algorithms, and leverage the mathematical framework of adaptive sub-modularity to obtain theoretical guarantees for the greedy-adaptive method.
\end{abstract}

\section{Introduction}
Lacking effective treatments or vaccinations, the most effective way to save lives in an ongoing epidemic is to mitigate and control its spread. 
This can be done by testing and isolating positive cases early enough to prevent subsequent infections. If done sufficiently regularly and for a sufficiently large fraction of individuals at risk, this has the potential to prevent a large fraction of the infections a positive case would normally cause.
However, a number of factors, such as limits on material resources as well as on work force, necessitate economical and efficient use of test resources.

\paragraph{Group testing}\footnote{\url{https://en.wikipedia.org/wiki/Group_testing}} aims to improve properties of tests by testing groups of items simultaneously.
We wish to leverage this framework to improve COVID-19 testing. One needs to differentiate between two different settings: \textbf{\textit{adaptive}} and \textbf{\textit{non-adaptive}}. In the former, tests can be decided one at a time, taking into account previous test results. In the latter, one has to select all tests before seeing any lab result and could thus run them in parallel. 

One can also imagine a \textbf{\textit{semi-adaptive}} setting in which tests are selected in small batches between each lab evaluation.
A simple example of a semi-adaptive group test is to first split $n$ samples into $g$ groups of (roughly) equal size, pool the samples within the groups and perform $g$ tests on the pooled samples. All samples in negatively tested pools are marked as negative, and all samples in positively tested pools are subsequently tested individually. This strategy has recently been validated for COVID-19 PCR tests \cite{Schmidt2020.04.28.20074187}.

\paragraph{Non-adaptive group testing.} Most of the existing research on non-adaptive group testing is concerned with identifying at most $k$ positive samples amongst $n$ items, which is referred to as non-adaptive {hypergeometric} group testing \cite{hwang1987non}. This assumption yields asymptotic bounds on the number of tests needed to recover the ground truth \cite{knill1998non,indyk2010efficiently,cheraghchi2012graph,chan2014non}. However, these are of limited practical relevance when constructive results on small numbers of samples are required. 
\paragraph{A different problem formulation.} We formulate the problem differently: given $n$ people and $m$ testing kits, the characteristics of the test and prior probabilities for each person to be sick, we seek to optimize the way the tests are used by combining several samples. For simplicity, samples are assumed to be independent. \cite{mazumdar2016nonadaptive} considers these assumptions as well, for the non-adaptive setting. However, his results are also asymptotic, \textit{i.e.}, valid for large $n$. 
\paragraph{Adaptive group testing.} In the adaptive setting, subsequent test designs may take into account earlier test results. By leveraging the framework of adaptive sub-modularity initially developed for sensor covering by \cite{golovin2011adaptive}, we prove near-optimality of the greedy-adaptive strategy.\\

\noindent\textbf{Our contributions} are as follows:
\begin{itemize}
    \item We provide a mathematically grounded objective function to optimize when designing a strategy, leveraging information theory. 
    \item We implement different strategies, both non-adaptive and adaptive, which can readily be used in a web browser to know \textit{(i)} which tests to run and \textit{(ii)} how to interpret the outcome. 
    \item We provide a guarantee of near-optimality of the greedy-adaptive strategy which is based on the mathematical objective we proposed. 
    \item Software is available at
\begin{center}
\url{https://louisabraham.github.io/crackovid/crackovid.html}.
\end{center}
\end{itemize}

\paragraph{Intuition.} \textit{Our mathematical objective is designed such that the mixture tests it proposes to run in the lab will maximize the amount of information we gain on the ground truth once their lab results are revealed} $-$ in expectation, over the randomness of both imperfect tests and prior probabilities of infection per individual.\\

\noindent\textbf{Notations} are progressively introduced throughout, but also all gathered in appendix~\ref{sec:notations}.

\section{Preliminaries}

Denote the number of patient\footnote{For simplicity, we will refer to all individuals being tested as {\em patients}.} samples by $n$, and the number of tests to run by $m$. Tests are assumed to be imperfect, with a {\em true positive rate (or sensitivity) $tpr$}\footnote{equivalent terms include {\em hit rate}, {\em detection rate} and {\em recall}.} and {\em true negative rate (or specificity) $tnr$}.\footnote{equivalent terms include {\em correct rejection rate} and {\em selectivity}.} patient sample $i$ is infected with probability $p_i\in [0,1]$ and we assume statistical independence of infection of patient samples. Denoting by a `1' a positive result (infection), the unknown ground truth describing who is infected and who is not is a vector of size $n$ made up of `0's and `1's: we call this the \textit{secret}, denoted as $s\in\{0,1\}^n$. A \textit{design of a test} $d\in\{0,1\}^n$ to run in the lab is a subset of patient samples to mix together into the same sample, where $d_i=1$ if patient sample $i$ is mixed into design $d$ and $d_i=0$ otherwise. Note that the outcome of a perfect design $d$ for a given secret $s$ can simply be obtained as $\mathbf{1}_{\langle d,s\rangle > 0}$ where $\langle d,s\rangle:=\sum_{i=1}^n d_i s_i$, i.e., a test result is positive if there is at least one patient $i$ for which $d_i=1$ ($i$ is included in the sample) and $s_i=1$ ($i$ is infected).

Recall that the secret $s$ is unknown.  However, since we assume initially that patient sample $i$ is infected with probability $p_i$ and that patient samples are independent,
this yields a \textit{prior} probability distribution over the possible values of $s$.  We hence represent the random value of $s$ as a \textit{random variable (r.v.)}, denoted by $S$, with probability distribution $p_S(s) := \Pr[S=s]$ over $\{0,1\}^n$.

Let us now recall the definition of the \textit{entropy} of our random variable,
\begin{equation}
    H(S) = -\sum_{s\in\{0,1\}^n} p_S(s)\log_2 p_S(s),
\end{equation}
representing \textit{the amount of uncertainty that we have on its outcome}, measured in bits. It is maximized when $S$ follows a uniform distribution, and minimized when $S$ constantly outputs the same value. As we perform tests, we gain additional knowledge about $S$. For instance, if a test pools all samples and returns a negative result, then our {\em posterior} probability that all patients are healthy goes up, i.e.,  $p_S((0,\dots,0))$ increases, governed by Bayes' rule of probability theory. More generally, we may perform a sequence of tests of varying composition, updating our posterior after each test. Our goal will be to select designs of tests so as to minimize entropy, resulting in the least amount of uncertainty about the test outcome for all individuals.

\section{Mathematical Framework}

Note that since tests are imperfect, for a given pool design $d\in\{0,1\}^n$ and a given secret $s\in\{0,1\}^n$, the Boolean outcome $T(s, d)$ of the test in the lab is not deterministic.
If tests were perfect, we would have $T(s,d)=\mathbf{1}_{\langle d,s\rangle > 0}$. To allow for imperfect tests, we model $T(s,d)$ as a r.v.\ whose distribution is described by
$\Pr[T(s,d) = 1 \mid \langle d,s\rangle > 0 ] = tpr$ and $\Pr[T(s,d) = 0 \mid \langle d,s\rangle = 0 ] = tnr$.\footnote{\label{footnote4}If we have prior information on whether and how these errors depend on how many samples are mixed into a given pool design (e.g., by dilution effects), we can take this into account by letting $tpr$ and $tnr$ depend on $|d|=\sum_i d_i$.} Since the secret $s$ is also unknown (and described by the r.v.\ $S$), the outcome $T(S,d)$ has now two sources of randomness: imperfection of tests and unknown secret.\footnote{Laboratory errors in composing the pooled designs $d$ could be modeled by correspondingly describing $d$ by a random variable, or by including these errors into the random variable $T$.}

In practice, one will not run one test but multiple tests. We now suppose that $m$ tests of pool designs are run and let their designs be represented as a multiset $\mathcal{D} \in (\{0, 1\}^n)^m$.\footnote{I propose to add this: It is a \textit{multiset} instead of just a \textit{set} because imperfection of lab tests allow strategies in which the same pool design may be tested multiple times.}

This leads us to the following question: given an initial prior probability distribution $p_S$ over the secret, how should we select pool designs to test in the lab? We want to select it such that once we have its outcome, we have as much information as possible about $S$, i.e. the entropy (uncertainty) of $S$ has been minimized. Since we cannot know in advance the outcome of the tests, we have to minimize this quantity \textit{in expectation} over the randomness coming from both the imperfects test and unknown secret. This requires the notion of \textit{conditional entropy}.

\paragraph{Conditional Entropy.} Given pool designs $\mathcal{D}$, we consider two random variables $S$ (secret) and $T:=T(S,\mathcal{D})$ (test results). The conditional entropy of $S$ given $T$ is given by:

\begin{equation}\label{eq:conditional}
H(S | T) = - \sum_{\substack{s\in\{0,1\}^n\\ t\in\{0,1\}^m}} \Pr[S=s, T=t] \cdot  \log_2\left(\frac{\Pr[S=s, T=t]}{\Pr[T=t]}\right) = \mathbb{E}_{t  \sim T(S, \mathcal{D})} \left[H(p_{S | T = t}) \right].
\end{equation}

In this formula, the joint probability $\Pr[S=s, T=t]$ has been computed with the conditional probability formula $\Pr[S=s, T=t] = \Pr[S=s] \Pr[T=t | S=s]$, and the posterior distribution is computed using Bayesian updating, i.e., 
\begin{equation}\label{eq:posterior}
    p_{S | T = t}(s) = \Pr[S=s | T=t] = \frac{\Pr[S=s, T=t]}{\Pr[T=t]},
\end{equation} where $\Pr[T=t] = \sum_s \Pr[S=s, T=t]$.

It represents the amount of information (measured in bits) needed to describe the outcome of $S$, given that the result of $T$ is known.

\paragraph{Mutual Information.} Equivalently, one can define the \textit{mutual information} between $S$ and $T$ as:
\begin{equation}
    I(S,T) := H(S) - H(S|T).
\end{equation}
It quantifies the amount of information obtained about $S$ by observing $T$.

\paragraph{A well-motivated criterion for test selection.} Since $H(S)$ does not depend on $d$, selecting the pool design $d$ minimizing the conditional entropy of $S$ given the outcome of $\mathcal{D}$ is equivalent to selecting the one maximizing the mutual information between $S$ and $T(S, \mathcal{D})$. We now have a clear criterion for selecting $\mathcal{D}$:
\begin{equation}\label{eq:criterion}
    \mathcal{D}^* \in \arg\max_\mathcal{D} I(S, T(S, \mathcal{D})).
\end{equation}
This criterion selects the pool designs $\mathcal{D}$ whose outcome will maximize our information about $S$.

\paragraph{Expected Confidence.} We report another evaluation metric of interest called the \textit{expected confidence}. It is the mean average precision of the maximum likelihood outcome. The maximum likelihood outcome it defined by:

\begin{equation}
    ML(t) := \arg\max_s \Pr[S=s | T=t],
\end{equation}
which yields the following definition of Expected Confidence:
\begin{align}
    \mathrm{Confidence}(S|T) :&= \Pr[S = ML(T)]\\
    &= \sum_{t\in\{0,1\}^m} \Pr[T=t, S=ML(t)]\\
    &=\mathbb{E}_{t  \sim T(S, \mathcal{D})} \left[\max_s p_{S | T=t}(s)\right]
\end{align}
$ML$ is of particular practical interest: given test results $t$, a physician wants to make a prediction. In this case, it makes sense to use the maximum likelihood predictor.
The interpretation of $\mathrm{Confidence}$ is straightforward: the probability of the prediction to be true (across all possible secrets).

\paragraph{Updating the priors.} Both scoring functions described above compute the expectation relative to the test results of a score on the posterior distribution $p_{S | T=t}(s)$. After observing the test results, we are able to replace the prior distribution $p_S$ by the posterior. By the rules of Bayesian computation, this update operation is commutative, i.e., the order in which designs $d_1$ and $d_2$ are tested does not matter, and compositional in the sense that we can test $\{d_1, d_2\}$ simultaneously with the same results.

Thus, we can decompose those steps and make different choices as we run tests (see the adaptive method below).

(A) Compare two outcomes: (1) the posterior is concentrated on two points $s\in \{0,1\}^n$, taking the values 0.6 and 0.4. (2) it is concentrated on three points, taking the values 0.6, 0.2, 0.2. The entropy is higher, but maybe we still prefer (2) since the margin to the second best explanation is larger?

\section{Non-Adaptive Method}
\paragraph{Foreword.} Non-adaptive methods have the benefit over their adaptive counterparts that they can be run in parallel. On the flip side, they describe a strictly more restrictive class of algorithms, since any non-adaptive method is an adaptive one ignoring the information obtained adaptively. Moreover, non-asymptotic (\textit{i.e.,} for small values of $n$) performance guarantees are harder to obtain than for adaptive methods. 

Given numbers $n$ \& $m$, test characteristics $tpr$ \& $tnr$ as well as prior probabilities of sample infection $p_i$, the best multiset $\mathcal{D}$ of $m$ pool designs is the one maximizing some score, like $I(S, T(S,\mathcal{D}))$ or $\mathrm{Confidence}(S, T(S,\mathcal{D}))$.

The tests are order insensitive, which gives a search space of cardinality ${2^n + m \choose m}$. Evaluating the score of every multiset separately takes $\mathcal{O}\left(2^{n+m}\right)$ operations.\footnote{We chose to implement a version with complexity $\mathcal{O}\left(m 2^{n+m}\right)$, but more cache efficient in practice.} Hence, brute-forcing this search space is prohibitive even for small values of $n$ and $m$.\\

We resort to randomized algorithms to find a good enough solution. Our approach is to use Evolution Strategies. We apply a variant of the $(1+\lambda)$ ES with optimal restarts \cite{luby1993optimal} to optimize any objective function over individuals (multisets of tests).

\paragraph{Detailed Description.} We maintain a population of $1$ individual between steps. At every step of the ES, we mutate it in $\lambda \in \mathbb{N}^+$ offsprings.
In the standard $(1+\lambda)$ ES, each offspring is mutated from the population, whereas our offsprings are iteratively mutated, each one being the mutation of the previous. These offsprings are added to the population, and the best element of the population is selected as the next generation of the population.

We initialize our population with the ``zero" individual that doesn't test anybody. Our mutation step is straightforward: flipping one bit $d_i$ of one pool design $d$, both chosen uniformly at random. Our iterative mutation scheme allows us to step out of local optima.

After choosing a basis $b$ proportional to $n \times m$ (which is approximately the logarithm of our search space), we apply restarts according to the Luby sequence: 
\begin{equation}\label{eq:restarts}
    (b,b,2 b,b,b,2 b,4 b,b,b,2 b,b,b,2 b,4 b,8 b,...)
\end{equation}

This sequence of restarts is optimal for Las Vegas algorithms \cite{luby1993optimal}, and our ES can be viewed as such under two conditions: \textit{(i)} that the population never be stuck in a local optimum, which can be achieved in our algorithm using $\lambda = n \times m$ (note that much smaller constant values are used in practice); \textit{(ii)} the second condition is purely conceptual and consists in defining a success as having a score larger than some (unknown) threshold. The fact that our algorithm does not use this threshold as an input yields the following result:

\begin{tcolorbox}
\begin{theorem}
Under condition \textit{(i)}, the evolutionary strategy using the sequence of restarts illustrated in Eq.~(\ref{eq:restarts}) yields a Las Vegas algorithm that restarts optimally (as defined by \cite{luby1993optimal}) to achieve any target score threshold.
\end{theorem}
\end{tcolorbox}

Future improvements would consist in \textit{(i)} initializing the population with random pool designs (following randomized testing methods from the literature), \textit{(ii)} design new mutation rules and  \textit{(iii)} a dynamic restart strategy based on the detection of lack of progress.

\section{Adaptive Method}

\paragraph{Foreword.} Adaptive methods have the clear advantage over their non-adaptive counterparts to be more efficient in the number of tests, although they require waiting for the lab results after each sequential test run. Although searching the space of all possible adaptive strategies would yield a prohibitive complexity of $\Omega(2^{2^m})$, it turns out that a simple adaptive strategy can yield provably near-optimal results.

\paragraph{Detailed Description.} We describe our adaptive method as Algorithm~\ref{alg:greedy} which greedily optimizes the criterion defined in Eq.~(\ref{eq:criterion}).

\begin{algorithm}[H]
\SetAlgoLined
\textbf{Input:} Numbers $n$ \& $m$, test characteristics $tpr$ \& $tnr$, priors $p_i$ for $i\in\{1,...,n\}$\;
\textbf{Output:} The sequence of tests to adaptively run in the lab\;
\textbf{Initialization:} Set $k:= m$ and set prior $p_S$ using the $p_i$'s\;
 \While{$k>0$}{
 For each pool design $d$ in $\{0,1\}^n$, compute $I(S,T(S, d))$\;
 Select any $d^*\in \arg\max_d I(S,T(S, d))$\;
 Observe result $T(S, d^*)$ of design $d^*$ in the lab\;
 Update $p_S$ accordingly (see Eq.~(\ref{eq:posterior}))
 to the realization of $d^*$ in the lab \;
 Decrease the number of remaining tests $k$ by $1$\;
 }
 \caption{(Greedy-Adaptive)}\label{alg:greedy}
\end{algorithm}
Leveraging the framework of adaptive sub-modularity \cite{golovin2011adaptive}, and the fact that the criterion defined by Eq.~(\ref{eq:criterion}) is adaptive sub-modular and adaptive monotone, Algorithm~\ref{alg:greedy} has the guarantee below. We prove it in Appendix~\ref{sec:proof-greedy}.
\begin{tcolorbox}
\begin{theorem}\label{thm:greedy} Denote by `$\mathrm{Algo}$' an adaptive strategy. Let $I(\mathrm{Algo})$ be the expected mutual information obtained at the end of all $m$ tests by running $\mathrm{Algo}$, the expectation being taken over all $2^m$ outcomes of lab results. Denote by `Optimal' the best (unknown) adaptive strategy. If we run Algorithm~\ref{alg:greedy} for $m_1$ tests and Optimal for $m_2$ tests, we have:
\begin{equation}
    I(\mathrm{Algorithm~\ref{alg:greedy}}) \geq \left(1-e^{-\frac{m_1}{\alpha m_2}}\right)I(\mathrm{Optimal}),
\end{equation}
where $\alpha$ is defined as follows: assume that our priors $p_i$ are wrong, in the sense that there exist constants $c,d$ with $c p_i\leq p_i'\leq d p_i$ for $i\in\{1,...,n\}$, with $c\leq 1$ and $d\geq 1$, where $p_i'$ denotes the true prior: we set $\alpha:=d/c$. 
\end{theorem}
\end{tcolorbox}
\paragraph{Remarks.} Theorem~\ref{thm:greedy} states that Algorithm~\ref{alg:greedy} is \textit{(i)} robust to wrong priors and \textit{(ii)} near-optimal in the sense that the ratio of its performance with that of the optimal strategy goes to $1$ exponentially fast in the ratio of the numbers of tests run in each algorithm. For $\alpha=1$ and $m_1=m_2$, this yields $1-1/e\simeq 0.63$.

\section{Additional Related Work}
There exists recent work applying group testing to COVID-19:  \cite{seifried2020pool} report using mini pools of patient samples of size 5 yielding no error in the prediction of healthy patients, over a set of 50 patient samples; \cite{evaluation2020yelin} report the possibility of mixing up to 32 patient samples together, with a false positive rate of 10\%; \cite{presstimesisrael} made a press release about the possibility of reliably testing mixtures including as many as 64 patient samples; \cite{kadri2020enhancing} mentions simple mathematical methods to approach the problem; \cite{deleforge2020maths} published a blogpost with appealing illustrations vulgarizing the mathematics of group testing.

\section{Conclusion \& Future Work}

We have discussed some interesting special cases using relatively restrictive assumptions. While theoretical results often require such assumptions, our implementation can in principle be generalized into various directions...

\paragraph{Different objective functions.} We have used the number of tests and samples as given, and then optimized a conditional entropy. However, from a practical point of view, other quantities are relevant and may need to be included in the objective, \textit{e.g.} the expectation (over a population) of the waiting time before an individual is ``cleared'' as negative (and can then go to work, visit a nursing home, or perform other actions which may require a confirmation of non-infectiousness). 

\paragraph{Semi-adaptive tests.} Instead of performing $m$ consecutive tests, one could do them in $k$ batches of respective sizes $m_1,...,m_k$ satisfying $m_1+...+m_k=m$.  Adaptivity over the sequence of length $k$ could be handled greedily as in Algorithm~\ref{alg:greedy}, except that instead of selecting a single pool design $d^*$, we would select $m_i$ designs at the $i^{th}$ step. We named this semi-adaptive algorithm the \textit{k-greedy} strategy.

\paragraph{Pool size optimization.} One could analyze the simple setting of \cite{Schmidt2020.04.28.20074187}, \textit{i.e.,} what is the best pool size given an average prior probability. ``Best'' here could mean something different than in our setting. One might want to take into account the number of tests kits used as well as the number of people which are ``cleared'' as negative after a given time duration.

\paragraph{Pool allocation.} How do we allocate people to the (first round) pool designs in this setting if prior probabilities are non-uniform? This will naturally arise if we apply group testing as per \cite{Schmidt2020.04.28.20074187} every day for part of the workforce of a company, in which case a negative result for a person on a given day would imply a low prior on the next day. One could also wonder how to allocate people to the pool designs in the presence of correlations between patient samples. One initial guess could be to place correlated people into the same pool in order to minimize the expected number of what one might call {\em false pool alarms}, i.e., cases where a person ends up in a positively tested pool even though they are personally negative. 

\paragraph{Further practical considerations.} A good practical strategy could be to perform one round of pooled tests to disjoint groups every morning as individuals arrive at work, being evaluated during work hours. Those who are in a positive group (adaptively) get assigned to a second pool design tested later, which can consist of a non-adaptive combination of multiple designs, tested over night. They receive the result in the morning before they go to work, and if individually positive, the enter quarantine. If the test is so sensitive that it detects infections even before individuals become contagious (which may be the case for PCR tests), such a strategy could avoid most infections at work.

\paragraph{Dependencies between $tpr$, $tnr$ and pool size.} The reliability of tests may vary with pool size. In our notation, the outcome of the tests is a random variable that need not only depend on whether one person is sick ($\mathbf{1}_{\langle d,s\rangle > 0}$) but it may also depend on the number of tested people $|d|$ and the number of sick people $\langle d,s\rangle$ (cf.\ Footnote~\ref{footnote4}); it could even assign different values of $tpr$ and $tnr$ to different people. The $tpr$ may in practice be an increasing function of the proportion of sick people ${\langle d,s\rangle} / {|d|}$.

\paragraph{Estimating prior infection probabilities.} Currently, we start with a factorized prior of infection that not only assumes independence between the tested patients but is also oblivious to the individual characteristics. We could, however, build a simple ML system that estimates the prior probabilities based on a set of features such as: job, number of people living in the same household, number of children, location of home, movement or contact data, etc.\footnote{subject to privacy considerations} Those prior probabilities can then be readily used by our approach to optimize the pool designs, and the ML system can gradually be improved as we gather more test results.

\paragraph{Prevalence estimation.}
Similar methods can be applied to the question of estimating prevalence. Note that this is an easier problem in the sense that we need not necessarily estimate which individuals are positive, but only how many.

\paragraph{Ethical considerations.}
We identify two families of concern to address. The first family concerns the accuracy of the tests. Indeed, when the number of tests and patients are equal, it is natural to compare the $tpr$/$tnr$ of the individual test to the $tpr$/$tnr$ of the individual results in our grouped test framework (obtained by marginalizing the posterior distribution). In some situations with unbalanced priors, the marginal $tpr$/$tnr$ of some people in the group could be lower than the test $tpr$/$tnr$, even if the test will be more successful overall. However, reporting the marginal individual results gives doctors a tool to decide whether further testing should be needed; hence we cannot rule out that individuals might be worse off by being tested in a group. 
\\
The second family of concerns, directly resulting from the first, is the responsibility of the doctor when assigning the people in batches and giving them prior probabilities (using another model). The assignment of people in batches should be dealt with in a future extension of our framework, while the sensitivity of our protocol to priors should be studied in more depth. In particular, the adaptive framework is more robust with respect to the choice of priors than the non-adaptive one.

\section*{Acknowledgments}
Gary B\'ecigneul is funded by the Max Planck ETH Center for Learning Systems.

\bibliographystyle{apalike}
\bibliography{biblio}

\newpage
\appendix

\section{List of All Notations}\label{sec:notations}
We use upper case letters exclusively for random variables (r.v.), except for mutual information $I$ and entropy $H$.
\begin{itemize}
    \item $n$: number of patient samples;
    \item $m$: number of tests to run in the lab;
    \item $s\in\{0,1\}^n$: the \textit{secret} to unveil, with $s_i=1$ if and only if patient sample $i$ is positive (infected);
    \item $S$: r.v. over possible values of $s$ whose law describes the current information we have about $s$;
    \item $d\in\{0,1\}^n$: a pool design, with $d_i=1$ if and only if patient sample $i$ belongs to pool design $d$;
    \item $\mathcal{D}\in(\{0,1\}^n)^m$: random multiset describing the pool designs output by the strategy;
    \item $t\in\{0,1\}^m$: lab result of a list of $m$ tests;
    \item $T$: r.v. over possible values of $t$ describing lab results;
    \item $tpr$: true positive rate, sensitivity, hit rate, detection rate, recall;
    \item $tnr$: true negative rate, specificity, correct rejection rate, selectivity;
    \item $p_i\in [0,1]$: prior probability of injection of patient sample $i$;
    \item $\mathrm{Pr}[A]$: probability of event $A$ to happen;
    \item $p_S(s)\in[0,1]$: probability of secret $s\in\{0,1\}^n$ to be the correct one, according to the law $p_S$ of r.v. $S$.
\end{itemize}

\section{Proofs}
\subsection{Theorem~\ref{thm:greedy}}\label{sec:proof-greedy}

We wish to invoke Theorem~1 of \cite{golovin2011adaptive}. In order to do so, we need to prove that the conditional entropy which we introduced in Eq.~(\ref{eq:conditional}) is both \textit{adaptive monotone} and \textit{adaptive sub-modular}. Direct respective correspondence between our notations and that of \cite{golovin2011adaptive} is given by:
\begin{itemize}
    \item Pool designs $d$ : items $e$;
    \item Test results $T$ : realizations $\Phi$;
    \item Set $\mathcal{D}$ of selected designs : set $E(\pi,\Phi)$ of selected items by policy $\pi$;
    \item $H(p_{S\mid T=t})$ : $f(E(\pi,\Phi), \Phi)$;
    \item $H(S\mid T)$ : $f_{avg}:=\mathbb{E}[f(E(\pi,\Phi), \Phi)]$.
\end{itemize}

This allows one to define, following Definition 1 of \cite{golovin2011adaptive}, the conditional expected marginal benefit of a pool design $d$ given results $t$ as:
\begin{equation}\label{eq:delta}
    \Delta(d):= - [H(S\mid R(S,d)) - H(S)].
\end{equation}
It represents the marginal gain of information obtained, in expectation, by observing the outcome of $d$ at a given stage (this stage being defined by $p_S$, i.e. after having observed test results $t$). \\

\textbf{Adaptive monotonicity} holds if $\Delta(d) \geq 0$ for any $d$.\\

\textbf{Adaptive sub-modularity} holds if for any two sets of results $t$ and $t'$ such that $t$ is a \textit{sub-realization}\footnote{\textit{i.e.} there exist $\mathcal{D}$ and $\mathcal{D}'$ such that $T(S,\mathcal{D})=t$, $T(S,\mathcal{D}')=t'$ and $\mathcal{D}\subset \mathcal{D}'$.} of $t'$, for any pool design $d$: $\Delta(d\mid t)\geq \Delta(d\mid t')$.\\

The below lemma concludes the proof.\\

\textbf{Lemma.} With respect to $\Delta$ defined in Eq.~(\ref{eq:delta}), adaptive monotonicity and adaptive sub-modularity both hold.\\
\textit{Proof.}
Adaptive monotonicity is a consequence of the ``information-never-hurts'' bound $H(X\mid Y)\leq H(X)$ \cite{cover2012elements}. Moreover, as mentioned in Lemma 1 of \cite{guestrin2005near}, sub-modularity also follows directly from this bound.

\begin{flushright}
$\square$
\end{flushright}

\section{Practical examples}
In this section, we show how our program (available at \url{https://louisabraham.github.io/crackovid/crackovid.html}) can be used for practical applications.

\subsection{Evaluation mode}

Suppose a doctor already made some pooled tests. Our program can compute the posterior distribution and report the most probable diagnosis, its confidence as well as the marginal probabilities for each person to be sick.

In our numerical example, \texttt{tpr=0.95, tnr=0.99}, we test 3 persons with 3 tests. The $i$-th test is applied to everybody but the $i$-th person. We can then enter the results of each test.

The input of our program is thus

\begin{verbatim}
3 3

0.99 0.95

0.1 0.1 0.1

eval 

011
101
110

[results]
\end{verbatim}

with \texttt{[results]} a binary code representing the test observations.

If the results are \texttt{000}, the program indicates that with very high probability nobody is sick:\footnote{To re-run in one click: \myurlevalmode}

\begin{verbatim}
most probable diagnosis: 000
confidence: 0.999963

marginals: 1.23414e-05 1.23414e-05 1.23414e-05 
\end{verbatim}

If the results are \texttt{011}, the program predicts that the first person is sick and the other two are healthy:

\begin{verbatim}
most probable diagnosis: 100
confidence: 0.973086

marginals: 0.975488 0.00292 0.00292 
\end{verbatim}

Interestingly, we observe error correction when the results are \texttt{001} (an impossible outcome with perfect tests) as the program still infers that nobody is sick:
\begin{verbatim}
most probable diagnosis: 000
confidence: 0.955646

marginals: 0.0221854 0.0221854 6.64093e-05 
\end{verbatim}

\subsection{Optimization mode}

This mode, given prior probabilities of people to be sick and a number of tests, finds a pooling scheme to optimize the expectation of some score on the posterior probabilities, eg minimize the entropy or maximize the confidence.

In fact, the framework we applied above is optimal as
\begin{verbatim}
3 3

0.99 0.95

0.1 0.1 0.1

optim confidence
ga-luby 2 100
1000
\end{verbatim}
outputs:\footnote{To re-run in one click: \myurloptimmode}

\begin{verbatim}
expected confidence:
0.958704

tests (one per line):
110
101
011
\end{verbatim}

Suppose that instead of testing 3 people, we want to test 6 people. A possible return value of our randomized algorithm is:

\begin{verbatim}
expected confidence:
0.937214

tests (one per line):
110100
100010
001110
101001
000101
010011

\end{verbatim}

As $0.958704^2 = 0.919113 < 0.937214$, grouping 6 people together gives much more accurate results than dividing then in two groups of 3 people.

\end{document}